\documentclass[aps]{revtex4}
\usepackage{graphicx,psfrag}
\def\bc{\begin{center}}
\def\ec{\end{center}}

\def\beq{\begin{equation}}
\def\eeq{\end{equation}}

\begin{document}

\title{
Anderson localization of two-dimensional Dirac fermions: \\ 
a perturbative approach
}

\author{K. Ziegler}
\affiliation{Institut f\"ur Physik, Universit\"at Augsburg\\
D-86135 Augsburg, Germany\\
}
\date{\today}

\begin{abstract}
Anderson localization is studied for two-dimensional Dirac fermions in the presence
of strong random scattering. Averaging with respect to the latter leads to a graphical 
representation of the correlation function with entangled random walks and three-vertices which 
connect three different types of propagators. This approach indicates Anderson localization along
a semi-infinite line, where the localization length is inversely proportional to the scattering rate.
\end{abstract}

\maketitle

\section{Introduction}

The physics of diffusion and Anderson localization is based on the picture that quantum 
particles scatter on impurities or defects of the underlying lattice structure. This 
represents a complex dynamical system which can be treated in practice only by some simplifying assumptions. 
First, we consider only independent particle of the system and average over all
possible scattering effects. For the latter we introduce a static distribution 
by assuming that the relevant scattering processes happen only on time scales 
that are large in comparison to the tunneling process of the quasiparticle in the lattice.

For the diffusive regime of such a disordered system exist powerful methods, such as
the nonlinear sigma model \cite{wegner79,wegner80,stone81} and the weak-localization approach 
\cite{gorkov79,hikami80,altshuler}. The latter is based on a perturbation
series in powers of $1/k_F l$, where $k_F$ is the Fermi wavector and $l$ the mean-free path.
This approach, however, is not directly applicable if $k_F\sim0$, which is, for instance, the
case for Dirac fermions at the spectral nodes \cite{andoetal,khveshchenko06,tkachov11,schmeltzer13}.
The special transport properties at these nodes have attracted great interest in the context of
graphene \cite{graphene1,graphene2} and topological insulators \cite{TI}. Therefore, it is important to develop
a flexible approach which allows us to study the related physics.

An alternative perturbative approach to the above mentioned 
methods was suggested recently, 
based on the idea that $E_b/\eta$ is a small parameter ($E_b$ is the band width, $\eta$ the scattering rate)
\cite{1404}.
The scattering rate is related to the scattering time $\tau$ by $\eta=\hbar/\tau$ and to the
mean-free path $l$ by $\eta=\hbar v_F/l$ ($v_F$ is the Fermi velocity). Thus, in contrast to the 
weak-localization approach the expansion parameter $E_b l/\hbar v_F$ depends on the bandwidth 
rather than on the Fermi wavevector $k_F$. This approach enables us to study the regime with
short mean-free path $l$, where we expect Anderson localization. The latter phenomenon is connected
with a special type of symmetry breaking: While diffusion breaks the time-reversal invariance
of the underlying microscopic system, Anderson localization breaks the scaling invariance of 
diffusion by creating a finite scale, the localization length. This is similar to the Kosterlitz-Thouless
transition in the XY model, where thermal fluctuations create vortex pairs whose correlation decays
exponentially \cite{glimm,itzykson}. This can be understood in a more formal way:
In the presence of weak disorder we have diffusion, characterized by the diffusion propagator 
$K_q\propto 1/(i\omega+Dq^2)$ with diffusion coefficient $D$, which has two poles 
$q_\pm=\pm \sqrt{-i\omega/D}$ for the wavevector $q$.
For Anderson localization, on the other hand, one would expect the appearence of poles away from the real axis,
where the distance from the latter is proportional to the inverse localization length.
However, this would imply that $K_{q=0}$ is finite in the limit $\omega\to0$, which
violates the general property $K_{q=0}\propto 1/i\omega$ \cite{wegner79a,mckane80}. It was found within
the strong-scattering expansion that there is only one pole that has a distance from the real
axis inversely proportional to the localization length, whereas the other pole approaches the
real axis with $\omega\to 0$ \cite{1404}. This result shall be used in this paper to study the localization
properties of 2D Dirac fermions in the presence of strong scattering.

The article is organized as follows. 
Fundamental quantities and the model are defined in Sect. \ref{sect:trans_prob}. Then in Sect. \ref{sect:pert_th} 
we briefly summarize the perturbation theory of Ref. \cite{1404}. The three progagators of the theory are discussed 
for 2D Dirac fermions in Sects. \ref{sect:props} and \ref{sect:g_discussion}. These results for the propagators
are used in Sect. \ref{sect:expansion} to show that the linked cluster expansion of the correlation function
is convergent for strong scattering. Finally, the results are summarized in Sect. \ref{sect:conclusions}.


\section{Transition probability}
\label{sect:trans_prob}

At weak scattering we expect
diffusion, a behavior known from classical physics, where the mean-square displacement of a particle position grows
linearly with time. This behavior is also valid for quantum systems \cite{thouless74}. It provides our basic understanding 
for a large number of transport phenomena, such as the metallic behavior in electronic systems.
Starting point is the transition probability for a particle, governed by the random Hamiltonian $H$,
to move from the site $r'$ on a lattice to another lattice site $r$ within the time $t$: 
\beq
P_{rr'}(t)=\sum_{j,j'}\langle|\langle r,j|e^{-iHt}|r',j'\rangle|^2\rangle_d
\ ,
\label{trans_prob}
\eeq
where $\langle ... \rangle_d$ is the average with respect to randomly distributed disorder.
The indices $j,j'$ refer to different bands of the system. In the following we will focus on the
specific case of the 2D Dirac Hamiltonian $H=v_F{\vec p}\cdot {\vec \sigma}+H_1$ where $H_1$ is 
a random term with mean zero, and where $v_F$ is the Fermi velocity. The components of the vector
${\vec \sigma}=(\sigma_1,\sigma_2)$ are Pauli matrices. Assuming a cut-off $\lambda$ for the momentum, there is
an effective bandwidth $E_b=2v_F\lambda^2$. In this case $j,j'=1,2$ are spinor indices.

With the expression (\ref{trans_prob}) we obtain, for instance, the mean-square displacement as
\[
\langle (r_k-r'_k)^2\rangle= \sum_r (r_k-r_k')^2 P_{rr'}(t)
\]
and the diffusion coefficient as
\[
D=\lim_{\epsilon\to 0}\epsilon^2\sum_r (r_k-r_k')^2 \int_0^\infty P_{rr'}(t) e^{-\epsilon t}dt
\ .
\]
The time integral in the last expression can also be written in terms of the Green's function as
\beq
\int_0^\infty P_{rr'}(t) e^{-\epsilon t}dt
=\int Tr_2\left\{G_{r,r'}(E+i\epsilon)\left[
G_{r',r}(E-i\epsilon) -G_{r',r}(E+i\epsilon)\right]\right\}dE
\label{msd1}
\ ,
\eeq
where $Tr_2$ is the trace with respect to the spinor index.
The one-particle Green's function
is defined as the resolvent $G(z)=(H-z)^{-1}$ of the Hamiltonian $H$, and $G_{r0}(E+i\epsilon)$ describes
the propagation of a particle with energy $E$ from the origin to a site $r$.

The correlation function of the Green's functions with poles on different half planes is dominant, 
whereas the correlation function of the Green's functions with poles only on one half plane is
the derivative of $Tr_2G_{r,r}(E+i\epsilon)$ with respect to $E$:
\[
\frac{\partial}{\partial E}Tr_2\left[G_{r,r}(E+i\epsilon)\right]
=\sum_{r'}Tr_2\left[G_{r,r'}(E+i\epsilon)G_{r',r}(E+i\epsilon)\right]
\ .
\]
Therefore, more important is the other term in (\ref{msd1}):
\beq
Tr_2\left[\langle G_{r,r'}(E+i\epsilon)G_{r',r}(E-i\epsilon)\rangle_d\right]
\ .
\eeq
A direct application of a perturbation theory for strong scattering would be an expansion of this
expression in powers of the off-diagonal terms $v_F{\vec p}\cdot {\vec \sigma}$ (hopping expansion). However, such an expansion
fails for small $\epsilon$ because the expansion terms diverge with $\epsilon\to0$. This is a consequence
of the fact that we have poles in the upper and in the lower half plane that move to the real axis for $\epsilon\to0$. 
It has been shown in Ref. \cite{1404} that this correlation of the Green's functions agrees 
for large distances $|r-r'|$ with the correlation function of a random-phase model, described by the expression 
\beq
K_{rr'}=\frac{\langle C^{-1}_{rr'}\det C\rangle_a}{\langle\det C\rangle_a}
\label{corr00}
\eeq
with
\beq
C_{rr'}= 2\delta_{rr'}-\sum_{j,j'}e^{i\alpha_{rj}}h_{rj,r'j'}\sum_{j'',r''}h^\dagger_{r'j',r''j''}e^{-i\alpha_{r''j''}}
\ .
\eeq
The brackets $\langle ...\rangle_a$ mean integration with respect to the angular variables 
$\{0\le\alpha_{rj}<2\pi\}$, normalized by $2\pi$. These angles represent the relevant part of the disorder fluctuations,
which are subject to long-range correlations of the Green's functions. 
Here it should be noticed that there is an invariance of $C$ with respect to a global phase change
$\alpha_{rj}\to \alpha_{rj}+\phi$.
Moreover, we have
\beq 
h_{rr'}=\sigma_0\delta_{rr'}+ 2i\eta(v_F{\vec p}\cdot {\vec \sigma}- i{\bar\eta})^{-1}_{rr'}\ \ \ {\rm with}\ {\bar\eta}=\eta+\epsilon
\ ,
\label{def_h}
\eeq
where 
$\eta\ge 0$ is the scattering rate in units of $v_F\lambda^2=E_b$.
In the limit $\epsilon\to 0$ the propagator $h$ is unitary:
\beq
hh^\dagger
={\bf 1}-4\epsilon(1-\epsilon){\bar\eta}(p^2+{\bar\eta}^2)^{-1}
\ .
\label{unitary0}
\eeq
It is convenient to introduce the generating functional
$
\log(\langle \det(C+a)\rangle_a)
$
with the $N\times N$ matrix $a$ ($N$ is the number of lattice sites). Then we obtain from (\ref{corr00})
\beq
K_{rr'}=\frac{\langle C^{-1}_{rr'}\det C\rangle_a}{\langle\det C\rangle_a}
=\frac{\partial}{\partial \alpha_{r'r}}\log(\langle \det(C+a)\rangle_a)\Big|_{a=0}
\ .
\label{corr01}
\eeq
In the remainder of this paper we will show that for strong scattering
\beq
K_{rr'}\sim \frac{1}{2} g_{r-r'}  
\ ,
\label{exp_decay}
\eeq
where $g_r$ decays exponentially, except for a line where it is constant.

\section{three-vertex expansion}

\subsection{General idea}
\label{sect:pert_th}

\begin{figure*}[t]
\includegraphics[width=9cm]{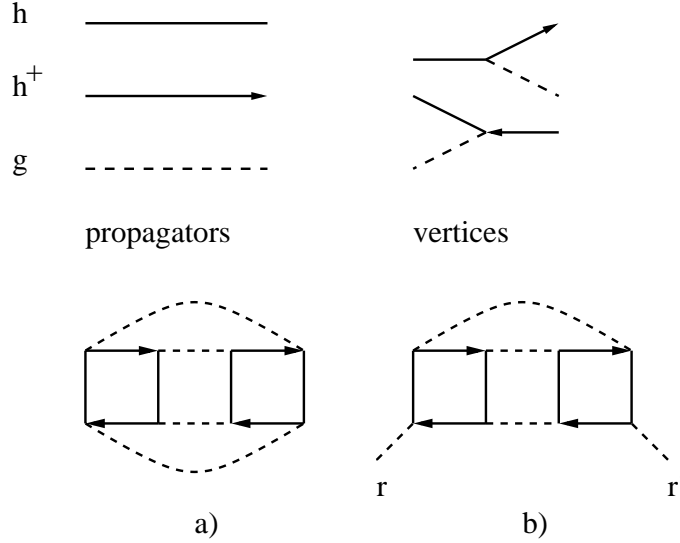}
\caption{
Propagators, vertices and two typical graphs of the linked cluster expansion:
a) is a graph of the generating function and b) is a graph of the correlation function $K_{rr'}$
of Eq. (\ref{corr01}), generated from graph a) by differentiation.
}
\label{fig:props}
\end{figure*}

We briefly recapitulate the perturbative expansion of Ref. \cite{1404}, which relies on the idea
that the expression (\ref{corr01}) can be treated within a linked cluster expansion of $\langle\det (C+a)\rangle_a$.
The latter is generated by the expansion of the determinant in Eq. (\ref{corr00})
\beq
\det(C+\alpha)=\frac{2^N}{\det g}e^A
\ \ \ {\rm with} \ \ A=Tr[\log({\bf 1}+\frac{1}{2}g(-C'+ {\bar C}))] \ \ \ {\rm and}\ 
g=\left({\bf 1}+\frac{1}{2}a -\frac{1}{2}{\bar C}\right)^{-1}
\label{det1}
\eeq
in powers of $\delta C=C'-{\bar C}$ around ${\bar C}$, where we use
\beq
C'_{rr'}=2\delta_{rr'}-C_{rr'}=\sum_{j,j'}e^{i\alpha_{rj}}h_{rj,r'j'}
\sum_{r'',j''}h^\dagger_{r'j',r'',j''}e^{-i\alpha_{r'',j''}}
\label{hop1}
\eeq
and 
\beq
{\bar C}_{rr'}=-\sum_{j,j'}h_{rj,r'j'}e^{i(\phi_j-\phi_{j'})}
\ .
\eeq 
For the last expression we have chosen fixed phases $\alpha_{rj}=\phi_j$ which are uniform in $r$. 
Then a Taylor expansion in the exponent of (\ref{det1}) yields 
\beq
A=Tr[\log({\bf 1}-\frac{1}{2}g\delta C)]=-\sum_{l\ge1}\frac{1}{2^l l}Tr\left[(g\delta C)^l\right]
\ .
\label{expansion3}
\eeq
This can be used to expand $e^A$ in powers of $\delta C$ and perform the angular integration for each expansion
term. The angular integration is easy to perform, since the expansion yields products of the phase
factors $e^{\pm i\alpha_rj}$, whose integration vanishes unless the phases compensate each other
in the product. 
Then the result of the angular integration has a graphical representation in terms of random walks, whose sites are connected pairwise
by the propagator $h^\dagger$. An equivalent representation consists of random walk whose steps are given by alternating propagators
$h$ and $h^\dagger$. The sites of these walks are connected pairwise by the propagator $g$.
This gives us eventually graphs that consist of three types of propagators, namely
$h$, $h^\dagger$ and $g$, and two types of three-vertices (cf. Fig. \ref{fig:props}).
Moreover, the linked cluster expansion is based on the relation $\langle e^A\rangle_a=e^{\langle A\rangle_c}$,
where $\langle A\rangle_c$ consists of those diagrams from the expansion of $\langle e^A\rangle_a$ which
are connected (or linked) graphs \cite{glimm,1404}. The latter provides an expansion, where the number
of terms increases exponentially with the number of propagators. Then we have a convergent expansion when
we can make the contribution of each propagator small. This will be discussed for 2D Dirac lattice fermions
with finite momentum cut-off $\lambda$ in the next section.  

\subsection{Propagators for 2D Dirac fermions}
\label{sect:props}

Now we have to analyze the three propagators of the three-vertex expansion for 2D Dirac fermions.
The three-vertices contribute only a factor $1$. 
The Fourier components of the propagators $h$ and $h^\dagger$ are
\beq
h_k\sim \frac{-1+\kappa^2}{1+\kappa^2}\sigma_0+2i\frac{{\vec\kappa}\cdot{\vec \sigma}}
{1+\kappa^2}, \ \ \
h^\dagger_k\sim \frac{-1+\kappa^2}{1+\kappa^2}\sigma_0-2i\frac{{\vec\kappa}\cdot{\vec \sigma}}
{1+\kappa^2} \ \ \ (\kappa=k/\eta)
\label{hs}
\eeq
for ${\bar\eta}\sim\eta$. These propagators decay exponentially in real space on the scale $\eta$, as explained
in App. \ref{app:prop_h}. This is important for the perturbation theory, since $\delta C$ contributes
at least one factor with an off-diagonal element of $h$:
\beq
\delta C_{rr'}=C'_{rr'}-{\bar C}_{rr'}=
\sum_{j,j'}e^{i\alpha_{rj}}h_{rj,r'j'}\sum_{r'',j''}h^\dagger_{r'j',r'',j''}e^{-i\alpha_{r''j''}}
+\sum_{j,j'}h_{rj,r'j'}e^{i(\phi_j-\phi_{j'})}
\ .
\eeq
In particular, the diagonal element reads
\beq
\delta C_{rr}
\sim -\sum_{j,j''}\sum_{r''\ne r}e^{i\alpha_{rj}}h^\dagger_{rj',r'',j''}e^{-i\alpha_{r''j''}}
\eeq
for $\epsilon\sim 0$ and $\eta\gg v_F\lambda^2$.
This expression contains only an off-diagonal propagator $h_{rr''}$ with $r''\ne r$. Therefore, 
$\delta C_{rr'}$ decays exponentially on the scale $1/\eta$. 
Thus, only the propagator $g$ determines the convergence of the perturbation series when we take $\eta\gg v_F\lambda^2$.
Its Fourier components are \cite{1404}
\beq
{\tilde g}_q=\frac{1}{1-\frac{1}{2}{\bar C}_q}
\sim\frac{\eta/2}{\epsilon +2i{\vec q}\cdot{\vec s}+4q^2/\eta}
\ ,
\label{prop_g1}
\eeq
where ${\vec s}=(\cos\Delta,\sin\Delta)$ depends on the global phase difference $\Delta=\phi_1-\phi_2$.
This propagator is more subtle than the propagators $h$ and $h^\dagger$.

\subsection{Discussion of the propagator $g$}
\label{sect:g_discussion}

The propagator ${\tilde g}_q$ is invariant under a global phase shift but it is sensitive to the difference
$\Delta$ of two uniform phases $\phi_1$ and $\phi_2$. In particular, the position of its poles with 
respect to $q_1$ depends on $c=\cos\Delta$, $s=\sin\Delta$:
\beq
q_\pm=-i\frac{\eta c}{4}\pm i\sqrt{\eta^2c^2/16 + \epsilon\eta/4+q_2^2+i\eta sq_2/2}
\ .
\label{poles}
\eeq
The corresponding poles with respect to $q_2$ are obtained by interchanging $c$ and $s$.
It should be noticed that any function of ${\tilde g}_q$ in which we integrate with respect to $q$
does not depend on $\Delta$. This is because ${\vec s}$ appears only as ${\vec q}\cdot{\vec s}=q\cos\varphi$,
where $\varphi$ is the angle between ${\vec q}$ and ${\vec s}$. An example of such a function is $\det g$.
In other words, the special choice of $\Delta$ affects only space-dependent quantities, such as the
correlation functions. On the other hand, fixing of $\Delta$ was only necessary to define a starting
point of our perturbation expansion. This indicates that $\Delta$ should be fixed by a variational
procedure to optimize the leading order. According to the standard procedure in mean-field
theories \cite{glimm,itzykson}, this would require a global quantity, such as a free energy.
The corresponding quantity in our case would be $\log(\det g)$, which, however, does not depend on $\Delta$.
Therefore, $\Delta$ plays a similar role as the phase angle in $U(1)$--symmetric models, such as
the XY model \cite{glimm,itzykson}. The propagator of the XY model does not depend on 
the phase angle, though. Thus, our theory, which is defined by the three propagators $h$, $h^\dagger$
and $g$, does not belong to any of the standard classes of field theory. The dependence of the
propagator $g$ on the choice of $\Delta$ is related to the fact that our system, defined by the
determinant $\det C$, is translational invariant, whereas Anderson localization
breaks scaling invariance by creating a finite length scale. This will be explained in more 
detail once the we have determined the behavior of $g_r$.

Before we Fourier transform ${\tilde g}_q$ of Eq. (\ref{prop_g1}) we consider the asymptotic case $\eta\sim\infty$,
where we neglect the quadratic term $4q^2/\eta$ in the denominator of ${\tilde g}_q$. Then the remaining 
linear term gives only one pole $q_+=i\epsilon/2c -(s/c) q_2$,
and the Cauchy integration yields
\beq
\frac{\eta}{2}\int_q\frac{e^{-iq\cdot r}}{\epsilon +2i{\vec q}\cdot{\vec s}}
=i\pi sgn(c)\eta \Theta(-cr_1) e^{\epsilon r_1/2c}\delta_{r_2,sr_1/c}
\eeq
with the Heaviside step function $\Theta$.
In the limit $\epsilon\to0$ this describes a propagator that vanishes everywhere except for the
line $r_2 
=\tan\Delta\Theta(-cr_1) r_1$. On this line its value is 
constant and proportional to the scattering rate.

\begin{figure*}[t]
\psfrag{r_1}{$r_1$}
\psfrag{r_2}{$r_2$}
\includegraphics[width=9cm]{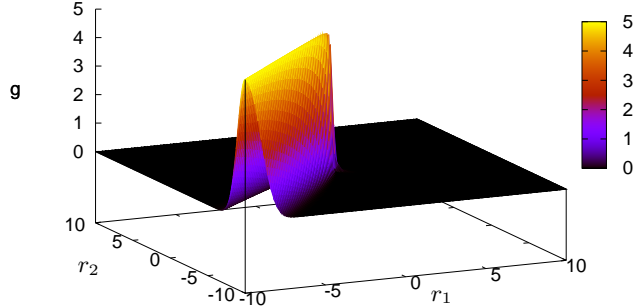}
\caption{
(Color online) 
The Propagator $g$  from Eq. (\ref{lcorr}) with $\Delta = \pi/4$ 
describes Anderson localization along a semi-infinite line.
}
\label{fig:prop_g}
\end{figure*}

The Fourier transform of the full propagator ${\tilde g}_q$ yields a very similar result, except for a softening
of the sharp line $r_2=\tan\Delta\Theta(-cr_1) r_1$: 
\beq
{\tilde g}_q\to g_r=\frac\eta{2}\int_q 
\frac{e^{-iq\cdot r}}{\epsilon +2i{\vec q}\cdot{\vec s}+4q^2/\eta}
\eeq
gives, according to App. \ref{app:prop_g}, an exponential decay of $g_r$ on the scale $1/\eta$
off the line $r_2= \tan\Delta\Theta(-cr_1) r_1$:
\beq
g_r\sim -\frac{\eta\pi}{8c}C e^{-\eta|cr_1|(s+|c|r_2/|r_1|)^2/8}\times\cases{
1 & for $cr_1<0$ \cr
e^{-\eta c r_1/2} & for $cr_1>0$ \cr
}
\ ,
\label{lcorr}
\eeq
where $\eta\gg v_F\lambda^2$, $\epsilon\sim 0$, and the coefficient $C$ is an integral given in Eq. (\ref{g_coeff}).
This propagator is depicted in Fig. \ref{fig:prop_g}.
Its behavior can be understood as Anderson localization away from this line.

\subsection{Convergent linked cluster expansion for strong scattering}
\label{sect:expansion}

The exponentially decaying behavior of $\delta C$ and of $g$ on the scale $1/\eta$ 
implies that the linked cluster expansion of the generating function
$
\log(\langle \det(C+a)\rangle_a)\Big|_{a=0}
$,
whose number of graphs grows exponentially with the number of vertices/propagators, is convergent
for sufficient large scattering rate $\eta$. The constant line of $g$ does not change this fact
because the $g$ appear in a loop, according to the Taylor expansion in Eq. (\ref{expansion3}):
Even if the constant line of $g$ contributes on a certain distance along the loop, this is only
possible in one direction. Then there is always a contribution from exponentially decaying terms
in the opposite direction, in order to close the loop. Therefore, the constant contribution is
compensated by an exponentially decaying contribution. 

Now we return to the right-hand side of Eq. (\ref{corr01}) to calculate the correlation function
$K_{rr'}$. There is the leading term from the prefactor in Eq. (\ref{det1})
\beq
-\frac{\partial \log(\det g)}{\partial \alpha_{r'r}}\Big|_{a=0}=\frac{1}{2}g_{rr'}
\ ,
\eeq
and for the expansion terms of the graphical representation the differentiation means breaking up a propagator $g$
into two propagators $g$, since only $g$ depends on $a$:
\beq
\frac{\partial g_{{\bar r}{\bar r}'}}{\partial \alpha_{r'r}}\Big|_{a=0}
=-\frac{1}{2}g_{{\bar r}r'}g_{r{\bar r}'}
\ .
\eeq
Thus, the correlation function is graphically a sum of random walks from $r$ to $r'$, as depicted in 
Fig. \ref{fig:props}b). These walks are estimated as $O(\eta^4 e^{-2\eta})$, where $\eta^2$ 
is the estimate of the two external $g$ propagators and $\eta^2e^{-2\eta}$ from the remaining loop.
The correlation function (\ref{corr01}) then is 
\beq
K_{rr'}=\frac{1}{2} g_{r-r'} + O(\eta^4 e^{-2\eta})
\ ,
\label{exp_decay2}
\eeq
which agrees with (\ref{exp_decay}).

\section{Conclusions}
\label{sect:conclusions}

Using the linked cluster expansion of Ref. \cite{1404}, we have studied the exponentially
decaying correlation function $K_{rr'}$ in (\ref{corr00}) for strong scattering. This expansion is constructed from three different
propagators, where two of them decay exponentially in all directions and one that decays exponentially 
only away from a semi-infinite line. Along this semi-infinite line it is constant. The three propagators
are connected by two types of vertices, as depicted in Fig. \ref{fig:props}. The leading term of the
convergent linked cluster expansion is given by the anisotropic propagator $g$ through (\ref{exp_decay2}).
Therefore, $g$ characterizes Anderson localization of 2D Dirac fermions at strong scattering. 
Here it should be noticed that we have considered a system on an infinite torus (i.e., for periodic boundary 
condictions in both directions). Other boundary conditions may change the location of the semi-infinite 
line, from which Anderson localization appears. In particular, this may also fix the direction of the
semi-infinite localization line, which is arbitrary on the infinite system.

\appendix

\section{Calculation of Propagators}
\label{app:prop}

\subsection{Propagator $h$}
\label{app:prop_h}

Fourier transformation of $h_k$ gives the $h_r$ that decays exponentially on the scaled $1/\eta$. 
\[
h_q\to h_r
=\int_q e^{-iq\cdot r}h_q
=\eta^2\int_q \frac{e^{-iq\cdot r}}{\eta^2+q^2}\left[\left(-1+\frac{q^2}{\eta^2}\right)
\sigma_0+\frac{2i}{\eta}{\vec q\label{app:prop_g}}\cdot{\vec \sigma}\right]
\]
\beq
=\sigma_0\int_q e^{-iq\cdot r}+2\eta^2\int_q \frac{e^{-iq\cdot r}}{\eta^2+q^2}
\left(-\sigma_0+\frac{i}{\eta}{\vec q}\cdot{\vec \sigma}\right)
=\sigma_0\delta_{r,0}+2\eta^2\int_q \frac{e^{-iq\cdot r}}{\eta^2+q^2}
\left(-\sigma_0+\frac{i}{\eta}{\vec q}\cdot{\vec \sigma}\right)
\ .
\label{h_r}
\eeq
The integral on the right-hand side is finite for an infinite cut-off. 
Therefore, we can perform an integration over the entire ${\bf R}^2$. 
This enables us to employ a Cauchy integration. Without restricting the generality we choose $r_j\ne0$
and obtain for $k\ne j$
\beq
I_{r}=\int_{-\lambda}^\lambda \int_{-\infty}^\infty \frac{e^{-iq_1r_1-iq_2r_2}}{\eta^2+q_1^2+q_2^2}dq_j dq_k
=\frac{\pi}{\eta}e^{-\eta |r_j|}\chi_k ,
\ \ \ \chi_k=\int_{-\lambda}^\lambda\frac{e^{-|r_j|\eta(\sqrt{1+q_k^2/\eta^2}-1)+iq_k r_k}}
{\sqrt{1+q_k^2/\eta^2}}dq_k
\ ,
\label{h_r2}
\eeq
with $|\chi_k|<\infty$. 
This yields
\beq
h_{r} = \sigma_0\delta_{r,0}-2\eta^2 I_{r}\sigma_0
- 2\eta\frac{\partial I_{r}}{\partial r_1}\sigma_1
- 2\eta\frac{\partial I_{r}}{\partial r_2}\sigma_2
\ ,
\eeq
whose off-diagonal terms decay exponentially on the scale $\eta$ according to (\ref{h_r2}). The diagonal term reads
\[
h_0=\sigma_0\left(1-2\eta^2\int_q\frac{1}{\eta^2+q^2}\right)\sim -\sigma_0\delta_{r,0}
\ ,
\]
where the asymptotic result is for $\eta\gg v_F\lambda^2$.

\subsection{Propagator $g$}
\label{app:prop_g}

We consider the expression in Eq. (\ref{lcorr}). Without restricting the generality we assume $r_j\ne 0$
and $k\ne j$. Then we perform the $q_j$ integration
first:
\[
g_r=\frac{\eta^2}{8}\int_{-\lambda}^\lambda\int_{-\infty}^\infty
\frac{e^{-iq_1 r_1-iq_2 r_2}}{\epsilon\eta/4+i\eta(q_1c+q_2s)+q_1^2+q_2^2}dq_jdq_k
=\frac{\eta^2}{8}\int_{-\lambda}^\lambda\int_{-\infty}^\infty
\frac{e^{-iq_1 r_1-iq_2 r_2}}{(q_+-q_j)(q_--q_j)}dq_jdq_k
\]
with $q_\pm$ defined in Eq. (\ref{poles}). A Cauchy integration gives
\beq
g_r=\frac{\eta^2 2i\pi}{8}\cases{
\int_{-\lambda}^\lambda\frac{e^{-iq_+r_1-iq_2r_2}}{q_--q_+}dq_2 & for $cr_1<0$ \cr
\int_{-\lambda}^\lambda\frac{e^{-iq_-r_1-iq_2r_2}}{q_--q_+}dq_2 & for $cr_1>0$ \cr
}
\ .
\eeq
This leads to
\beq
g_{r}=\Gamma_2\cases{
1 & for $cr_1<0$ \cr
e^{-\eta c r_1/2} & for $cr_1>0$ \cr
}
\ ,
\eeq
where
\beq
\Gamma_2=-\frac{\eta\pi}{8c}\int_{-\lambda}^\lambda
\frac{e^{-\eta (\sqrt{1+4\epsilon/\eta c^2+16 q_2^2/\eta^2c^2+i8s q_2/\eta c^2}-1)|cr_1|/4-iq_2r_2}}
{\sqrt{1+4\epsilon/\eta c^2+16 q_2^2/\eta^2c^2+i8s q_2/\eta c^2}}dq_2
\eeq
and $\Gamma_1$ after exchanging $c$ and $s$. Thus, we have $|\Gamma_k|<\infty$. 
Moreover, we expand the exponent and the denominator for $\eta\gg v_F\lambda^2$. In leading order we obtain
\beq
\Gamma_2\sim -\frac{\eta\pi}{8c}\delta_{s|cr_1|/c^2,-r_2} 
\ ,
\eeq
and when we also include terms with $1/\eta$ in the exponent
\beq
\Gamma_2\sim -\frac{\eta\pi}{8c}e^{-\eta|cr_1|(s+|c|r_2/|r_1|)^2/8}\int_{-\lambda}^\lambda e^{-|cr_1|q_2^2/\eta c^4}dq_2
\ .
\label{g_coeff}
\eeq

\end{document}